\documentclass{aa}  

\usepackage{graphicx}
\usepackage{txfonts}
\usepackage[colorlinks=true, allcolors=blue]{hyperref}
\usepackage{multirow}
\usepackage{dirtytalk}
\usepackage{xcolor}
\usepackage[normalem]{ulem}
\usepackage{longtable}

\titlerunning{Signatures of jets and winds in the OIR spectrum of GX 339$-$4}

\begin{document}

   \title{State-dependent signatures of jets and winds in the optical and infrared spectrum of the black hole transient GX 339$-$4}

   \subtitle{}

   \author{A. Ambrifi 
          \inst{1,}\inst{2}\fnmsep\thanks{ambrifi.astronomy@gmail.com}
          \and
          D. Mata Sánchez\inst{1,}\inst{2}
          \and
          T. Muñoz-Darias\inst{1,}\inst{2}
          \and
          J. Sánchez-Sierras\inst{1,}\inst{2}
          \and
          M. Armas Padilla\inst{1,}\inst{2}
          \and
          M. C. Baglio\inst{3,}\inst{4}
          \and
          J. Casares\inst{1,}\inst{2}
          \and
          J. M. Corral-Santana\inst{5}
          \and
          V. A. Cúneo\inst{6}
          \and
          R. P. Fender \inst{7,}\inst{8}
          \and 
          G. Ponti\inst{4,}\inst{9}
          \and
          D. M. Russell\inst{3}
          \and
          M. Shidatsu\inst{10}
          \and
          D. Steeghs\inst{11}
          \and
          M. A. P. Torres\inst{1,}\inst{2}
          \and
          Y. Ueda\inst{12}
          \and
          F. Vincentelli\inst{13}
          }

   \institute{Instituto de Astrofísica de Canarias, E-38205 La Laguna, Tenerife, Spain
         \and
        Departamento de astrofísica, Universidad de La Laguna, E-38206 La Laguna, Tenerife, Spain
        \and
        Center for Astro, Particle, and Planetary Physics, New York University Abu Dhabi, P.O. Box 129188, Abu Dhabi, UAE
        \and
        INAF–Osservatorio Astronomico di Brera, Via Bianchi 46, I-23807 Merate (LC), Italy
        \and
        European Southern Observatory, Alonso de Córdova 3107, Vitacura, Casilla 19001, Santiago de Chile, Chile
        \and
        Leibniz-Institut für Astrophysik Potsdam (AIP), An der Sternwarte 16, 14482 Potsdam, Germany
        \and
        Astrophysics, Department of Physics, University of Oxford, Keble Road, Oxford OX1 3RH
        \and
        Department of Astronomy, University of Cape Town, Private Bag X3, Rondebosch 7701, South Africa
        \and
        Max-Planck-Institut für extraterrestrische Physik, Giessen-
        bachstra{\ss}e 1, 85748 Garching bei München, Germany
        \and
        Department of Physics, Ehime University, 2-5, Bunkyocho, Matsuyama, Ehime 790-8577, Japan
        \and
        Department of Physics, University of Warwick, Gibbet Hill Road, Coventry CV4 7AL, UK
        \and      Department of Astronomy, Kyoto University, Kitashirakawa-Oiwake-cho, Sakyo-ku, Kyoto, Kyoto 606-8502, Japan
        \and
        Department of Physics \& Astronomy. University of Southampton,
Southampton SO17 1BJ, UK
        }

   \date{Received Month day, year; accepted Month day, year}

  \abstract{GX 339$-$4 is one of the prototypical black hole X-ray transients, exhibiting recurrent outbursts that allow detailed studies of black hole accretion and ejection phenomena. In this work we present four epochs of optical and near-infrared spectroscopy obtained with X-shooter at the Very Large Telescope. The dataset includes two hard state spectra, collected during the 2013 and 2015 outbursts, and two soft state spectra observed during the 2021 outburst. Strong Balmer, Paschen, \ion{He}{i} and \ion{He}{ii} emission lines are consistently observed in all spectra, while Brackett transitions and the Bowen blend are only prominent in the soft state. Although P-Cygni profiles are not identified, the presence of wind signatures, such as extended emission wings, flat-top and asymmetric red-skewed profiles, is consistently observed through most emission lines, suggesting the presence of wind-type ejecta. These features are particularly evident in the hard state, but they are also observed in the soft state, especially in the near-infrared. This strengthens the case for state-independent winds in black hole transients and increases the evidence for wind signatures in low-to-intermediate orbital inclination systems. We also study the spectral energy distribution, which provides evidence for the presence of synchrotron emission during the hard state. The jet significantly affects the near-infrared continuum, greatly diluting the emission features produced in the accretion flow. The simultaneous identification of both jet and wind signatures during the hard state reinforces the idea of a complex outflow scenario, in which different types of ejecta coexist.}

    \keywords{Accretion, accretion discs – X-rays: binaries – Stars: black holes – Stars: winds, outflows –  Stars: individual (V821 Arae) }

     \maketitle
%
%-------------------------------------------------------------------

\section{Introduction}

Most black hole transients (BHTs) are low mass X-ray binaries showing recurrent outbursts (e.g. \citealt{McClintock2006, Belloni2011, CorralSantana2016}). 
During these events their X-ray luminosity increases by several orders of magnitude and mass ejecta occur, both in the form of winds and jets (see, e.g. \citealt{Fender2016}). 

Compact, relativistic jets emitting synchrotron radiation (e.g. \citealt{BlandfordKonigl1979}) are detected during the hard state of the outburst, when a hard power-law characterises the X-ray spectrum (e.g. \citealt{Gallo2003}). These jets dominate the radio band and can be detected up to the infrared and optical regimes (e.g. \citealt{Corbel2001, Fender2001}). The jet is quenched during the soft state (e.g. \citealt{Russell2011}), which is characterised by a soft, thermal X-ray spectrum. Conversely, accretion disc winds are primarily observed in the X-ray band during this state (e.g. \citealt{Neilsen2009, Ueda2009, Ponti2012, Parra2024}). These hot winds are associated with high values of the ionisation parameter ($\xi_\mathrm{X}\geq10^3$ erg cm s$^{-1}$; e.g. \citealt{DiazTrigo2016}), and densities in the range of $n\sim10^{13}-10^{15}$ cm$^{-3}$ have been proposed (e.g. \citealt{Schulz2008, Kallman2009}; see also \citealt{Kosec2024} for a theoretical study).

\begin{table*}
\caption{Journal of observations.}             
    \label{table_obs}      
    \centerline{        
    \begin{tabular}{c c c c c c c c c c c}     % 7 columns 
    \hline                           
    Epoch & Date / MJD start & Orbital phase\tablefoottext{a} & Arm & Slit width (") & R ($\lambda/\Delta\lambda$) & \# & T\textsubscript{exp}(s) \\ 
    \hline  
        &  & & UVB & 1.0 & 5400 & 2 & 1162 \\
      S2021a & 11 Apr 2021 / 59315.35 & 0.5$\pm$0.3 & VIS & 0.9 & 8900 & 2 & 1135  \\
      &  & & NIR & 0.9 & 5600 & 4 & 300  \\
     \hline
        &  & & UVB & 1.0 & 5400 & 2 & 1162  \\
       S2021b & 15 Apr 2021 / 59319.27 & 0.7$\pm$0.3 & VIS & 0.9 & 8900 & 2 & 1135  \\
        & & & NIR & 0.9 & 5600 & 4 & 300  \\
       \hline
        & & & UVB & 1.0 & 5400 & 16 & 150 \\  
        H2013 & 16 Oct 2013 / 56581.00 & 0.7$\pm$0.2 & VIS & 0.9 & 8900 & 16 & 150 \\  
        & & & NIR & 0.9 & 5600 & 32 & 75 \\  
       \hline
        & & & UVB & 1.3 & 4100 & 12 & 142 \\
        H2015 & 01 Oct 2015 / 57296.01 & 0.30$\pm$0.05 & VIS & 1.2 & 6500 & 12 & 150 \\
        & & & NIR & 1.2 & 4300 & 36 & 43 \\
    \hline                  
    \end{tabular}}
    \tablefoot{
    \tablefoottext{a}{Orbital phases and their uncertainties were calculated using the ephemerides in \citeauthor{Heida2017} (\citeyear{Heida2017}).}}
\end{table*}
    
Over the past few years, a new ingredient has been added to the above picture: the detection of cold (i.e. low-ionisation) winds in several BHTs (e.g. \citealt{Munoz2016, MataSanchez2018}). This is inferred from the presence of conspicuous P-Cygni line profiles, blue-shifted absorption troughs, broad emission line wings and asymmetries in optical and near-infrared (NIR) emission lines. Optical signatures of wind-type outflows have been primarily detected during the hard state, sometimes simultaneously with the jet, in prominent emission lines, such as \ion{H$\rm{\alpha}$}{}, \ion{He}{i} line at $5876$\AA$ $ (\ion{He}{i}-$5876$ hereafter) and \ion{He}{i}-$6678$ (e.g. \citealt{Munoz2018, MunozDarias2019, Cuneo2020, MataSanchez2022}). These winds are characterised by a much lower ionisation parameter ($\xi_\mathrm{V}\lesssim10^{-3}\xi_\mathrm{X}$; see, e.g. \citealt{Munoz2022}), while numerical simulations suggest that their densities are comparable to those of hot winds ($n\sim10^{13}$ cm$^{-3}$; \citealt{Koljonen2023}).

In the NIR, evidence for winds is limited to a few systems. Wind signatures are mainly observed in emission lines of the Paschen and Brackett series (Pa$\rm{\beta}$, Pa$\rm{\gamma}$, Br$\rm{\gamma}$), as well as in \ion{He}{i}-$10830$. NIR winds have been detected during both the hard and soft states, pointing to a scenario in which wind-type ejecta are present throughout most of the outburst (e.g. \citealt{sanchezsierras2020}; \citealt{Panizo2022}; \citealt{Sanchezsierras23_1915}).

Interestingly, both hot and cold winds have been detected almost exclusively in intermediate to high
inclination systems, leading to the commonly accepted idea that
these outflows have a preferentially equatorial geometry (\citealt{Ponti2012, Panizo2022}). The detection of wind signatures at different wavelengths and phases of the outburst (see, e.g. \citealt{CastroSegura2022, Fijma2023, Sanchezsierras23_1915}), suggests that the detectability of the winds is also influenced by the physical properties of the ejecta. This points towards a multi-phase nature of the winds, which is supported by the similarity of their kinetic properties in simultaneous X-ray and optical data \citep{Munoz2022}.

In this paper we present optical and NIR spectroscopy of the BHT GX 339$-$4. This system was first discovered in 1973 (\citealt{Markert1973}) and has since shown frequent outbursts, with a recurrence period of $\sim2.5$ years. It hosts a K-type sub-giant star feeding a $M_{\rm{BH}}\approx 4-11 M_\odot$  black hole with an orbital period of $P_{\rm{orb}} = 1.76$ d (\citealt{Hynes2003, Munoz2008, Heida2017, Zdziarski2019}). Its orbital inclination is thought to fall within the range $40^\circ\lesssim i \lesssim 60^\circ$ \citep{Zdziarski2019}, making GX 339$-$4 a low or intermediate inclination system (see also \citealt{Munoz2013}). Previous spectroscopic studies of this source at limited spectral resolution revealed that its optical and NIR spectrum is dominated by emission lines of low ionisation elements (\ion{H}{I}, \ion{He}{I}, \ion{He}{II}). Single-peaked, round-top \ion{H$\rm{\alpha}$}{} profiles were observed in the hard state optical spectra presented by \citealt{Soria1999} and \citealt{Wu2001}, and were interpreted as a possible signature of a dense outflow. This conclusion is further supported by \citealt{Rahoui2014}, who also detected broad \ion{Pa$\rm{\beta}$}{} line wings during the hard state.

In January 2021, GX 339$-$4 entered into a new outburst (e.g. \citealt{Garcia2021, Sguera2021}). Here, we analyse two epochs of optical and NIR spectroscopy collected during the soft state phase. We also consider two archival spectra obtained during the hard state of two previous outbursts (2013 and 2015) to study the influence of the X-ray state in the optical and NIR spectral properties. These simultaneous optical and NIR spectroscopic observations represent the highest-resolution outburst data on this source currently available in the literature. The high signal-to-noise ratio of our data enables us to investigate outflow-related signatures across a significant number of emission lines during both the hard and soft states.

\begin{figure}
    \centering
    \includegraphics[width=0.5\textwidth]{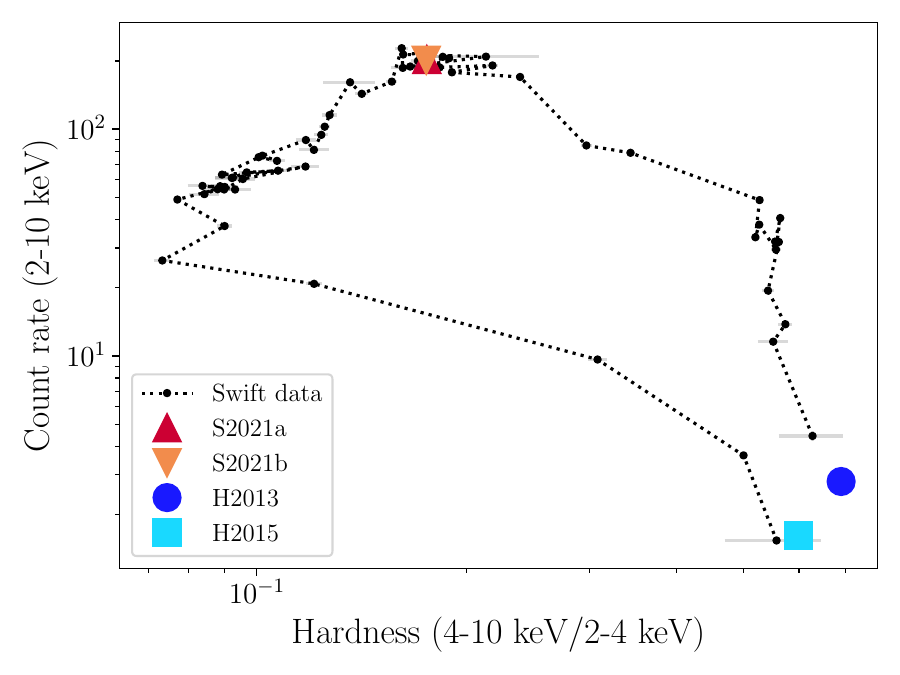}
    \caption{Swift-XRT HID during the 2021 outburst of GX 339$-$4. The epochs S2021a and S2021b are marked with up- and down-pointing triangles (overlapping in the figure), correspondingly. The hard state epochs, H2013 and H2015, are marked with a circle and a square, respectively.}
    \label{HID}%
\end{figure}

\section{Observations and data reduction}

Four epochs of spectroscopy were obtained with the X-shooter spectrograph (\citealt{Vernet2011}) at the Very Large Telescope (VLT) in Cerro Paranal, Chile. The instrument consists of three spectroscopic arms, operating simultaneously in the UVB (3000-5600 \AA), VIS (5500-10200 \AA) and NIR (10200-24800 \AA) spectral ranges. Two observations were performed on April 11 (MJD 59315) and April 15 (MJD 59319) 2021 (hereafter S2021a and S2021b, respectively). We also analysed two X-shooter archival spectra of the same source, which were collected during the 2013 and 2015 outbursts (hereafter H2013 and H2015). For each observation, 8 to 64 exposures in nodding configuration were obtained, with individual exposure times in the range of $\sim 40-1200$ s and resolutions of $\sim30-70$ km/s. In order to optimize sky subtraction, the source was observed using two nod positions. The details of the observations are provided in Table \ref{table_obs}. The absolute orbital phase for each epoch, based on the ephemeris by \citet{Heida2017}, is also reported in the table.  

The data were reduced using the X-shooter Pipeline version 3.6.1 (\citealt{Freudling2013}), which provides sky-subtracted, wavelength-calibrated mono-dimensional spectra. For each observing epoch, flux calibration was performed using a standard star observed close in time, either on the same or preceding night. In order to improve the flux calibration throughout the entire band covered by the spectra, a correction routine was implemented to compensate for the effects of wavelength-dependent light loss due to variable atmospheric seeing. To achieve this, a Gaussian was fitted at each wavelength to the spatial profile of the trace in the rectified and merged 2D spectra. This process was repeated for both the source and the standard star, effectively producing wavelength-dependent solutions for the seeing. We calculated the integrated area of the profiles within the region truncated by the fixed slit width ($A_{ob,\lambda}$ for the source and $A_{std,\lambda}$ for the standard star). This allowed us to derive a correction factor for the flux lost due to slit truncation, defined as $f_{\lambda}=A_{std,\lambda}/A_{ob,\lambda}$, which was then multiplied by our target spectra. We find $f_{\lambda}$ values close to unity, with a median of $\sim0.9$. As expected, the points that significantly deviate are those at the joints between the different X-shooter arms.

All the spectra are Doppler-shifted to the barycentric rest frame. \texttt{MOLECFIT} (\citealt{Smette2015, Kausch2015}, version 1.5.9) was used for correcting telluric absorption in the optical and NIR ranges.  The spectra were dereddened using the \texttt{dust\_extinction} python package (\citealt{Gordon2022}). The dereddening was performed using the \texttt{F19} routine, which is based on the model by \cite{Fitzpatrick2019}. We adopted the extinction law $A_{\rm{V}}=3.1 E(B-V)$ by \citeauthor{Cardelli1989} (\citeyear{Cardelli1989}),  along with a previously reported colour excess $E(B-V) = 1.2\pm0.1$ (\citealt{Zdziarski1998}).

Finally, we note that our analysis is complemented by daily averaged Swift-XRT data obtained in Windowed Timing mode. We used count rates in the 2-10 keV range and hardness ratios (4-10 keV / 2-4 keV) corresponding to the 2013, 2015, and 2021 outbursts. These data were used solely to approximately trace the X-ray behaviour of the source at the time of our spectroscopic observations (see below) and were directly derived from the Swift-XRT data products generator\footnote{\url{https://www.swift.ac.uk/user_objects/}} (\citealt{Evans2007}, \citeyear{Evans2009}). The data analysis presented in the next sections was performed using custom routines developed under Python 3.9.

\section{Analysis and results}

The evolution of the 2021 outburst across the hardness-intensity diagram (HID; e.g. \citealt{Homan2001}),  is shown in Fig. \ref{HID} (see also \citealt{Stiele2023}). The outburst transitioned from the hard to the soft state and returned to quiescence following the typical hysteresis pattern of BHTs (e.g. \citealt{Fender2004}). The approximate location of our spectroscopic observations is marked in the diagram. We note that the spectra S2021a and S2021b were collected during the most X-ray luminous stages of the outburst. The position of the spectroscopic epochs H2013 and H2015 in the HID was determined using data from their corresponding outbursts. In particular, H2013 was collected during the decay of a failed outburst (\citealt{Belloni2013}), whereas H2015 was observed during the final stages of the 2015 complete outburst. Their location in the HID diagram confirms the classification of S2021a and S2021b as soft state (\citealt{Stiele2023, Liu2023}), and H2013 and H2015 as hard state observations (\citealt{Belloni2013, Tomsick2015}).

\subsection{Evolution of the continuum}
\begin{figure}
    \centering
    \includegraphics[width=0.5\textwidth]{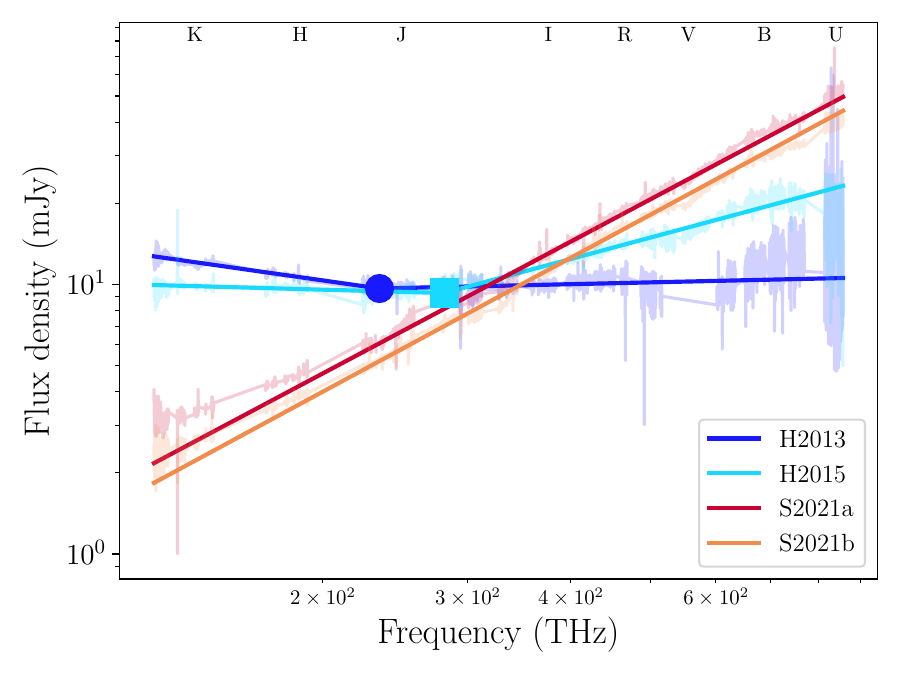}
    \caption{Fit to the dereddened continuum of the four epochs. The fit excludes regions with strong telluric absorptions. The symbols, coded as in the previous figures, mark the break location. }
    \label{continuum}%
\end{figure}

\begin{table}
\caption{Best-fit indices obtained from fitting the optical to NIR SEDs.} 
\label{table_best_fit_cont}      
\centerline{        
\begin{tabular}{c c c c}     
    \hline                           
    Epoch & $\nu_0$ (THz) & $\rm{\alpha_{NIR}}$ & $\rm{\alpha_{OPT}}$ \\ 
    \hline  
    S2021a\tablefoottext{*} & - & $1.6\pm0.3$& $1.6\pm0.3$ \\
    S2021b\tablefoottext{*} & - & $1.7\pm0.3$& $1.7\pm0.3$ \\
    H2013\tablefoottext{\#} & $235\pm1$ & $-0.43\pm0.08$&  $0.1\pm0.2$\\
    H2015\tablefoottext{\#} & $282\pm1$ & $-0.08\pm0.08$&  $0.8\pm0.3$\\
\hline                  
\end{tabular}}
\tablefoot{\tablefoottext{*}{Soft state epochs were fitted with a single power-law (i.e. $\rm{\alpha_{NIR}} = \rm{\alpha_{OPT}}$)}. 
\tablefoottext{\#}{Hard state epochs were fitted with a broken power-law.} }
    \end{table}

The optical to NIR spectral energy distribution (SED) of the four epochs is presented in Fig. \ref{continuum}. The spectral continuum of the two soft state epochs (S2021a and S2021b) is almost identical. Likewise, the SED in the two hard state epochs is similar, despite corresponding to different outbursts. We note that the hard state spectra exhibit an enhanced NIR continuum when compared to those of the soft state.  To further investigate this, the hard state SEDs were fitted with a broken power law given by:
\begin{equation}
f_{\rm{\nu}}(\rm{\nu}) = 
\begin{cases}
      a\cdot\big(\frac{\rm{\nu}}{\nu_0}\big)^{\rm{\alpha_{OPT}}} & \nu\geq\nu_0\\
      a\cdot\big(\frac{\rm{\nu}}{\nu_0}\big)^{\rm{\alpha_{NIR}}} & \nu<\nu_0,
    \end{cases}
\end{equation}
where $a$, $\nu_0$, $\alpha_{\mathrm{OPT}}$ and $\alpha_{\mathrm{NIR}}$ are free parameters. Conversely, a single power law was employed for fitting the soft state SEDs (i.e. $\alpha=\rm\alpha_{\rm{OPT}}=\rm\alpha_{\rm{NIR}})$. The fit excludes regions heavily affected by telluric absorption and diffuse interstellar bands (DIBs). Emission lines have also been masked. The region 520-600 THz was excluded from the fit to the H2013 SED due to a significant flux drop at the frequency connection between the UVB and VIS X-shooter arms (possibly due to the disabling of the atmospheric dispersion correctors of the instrument during this epoch). 

\begin{figure*}
    \centering
    {\includegraphics[width=\textwidth]{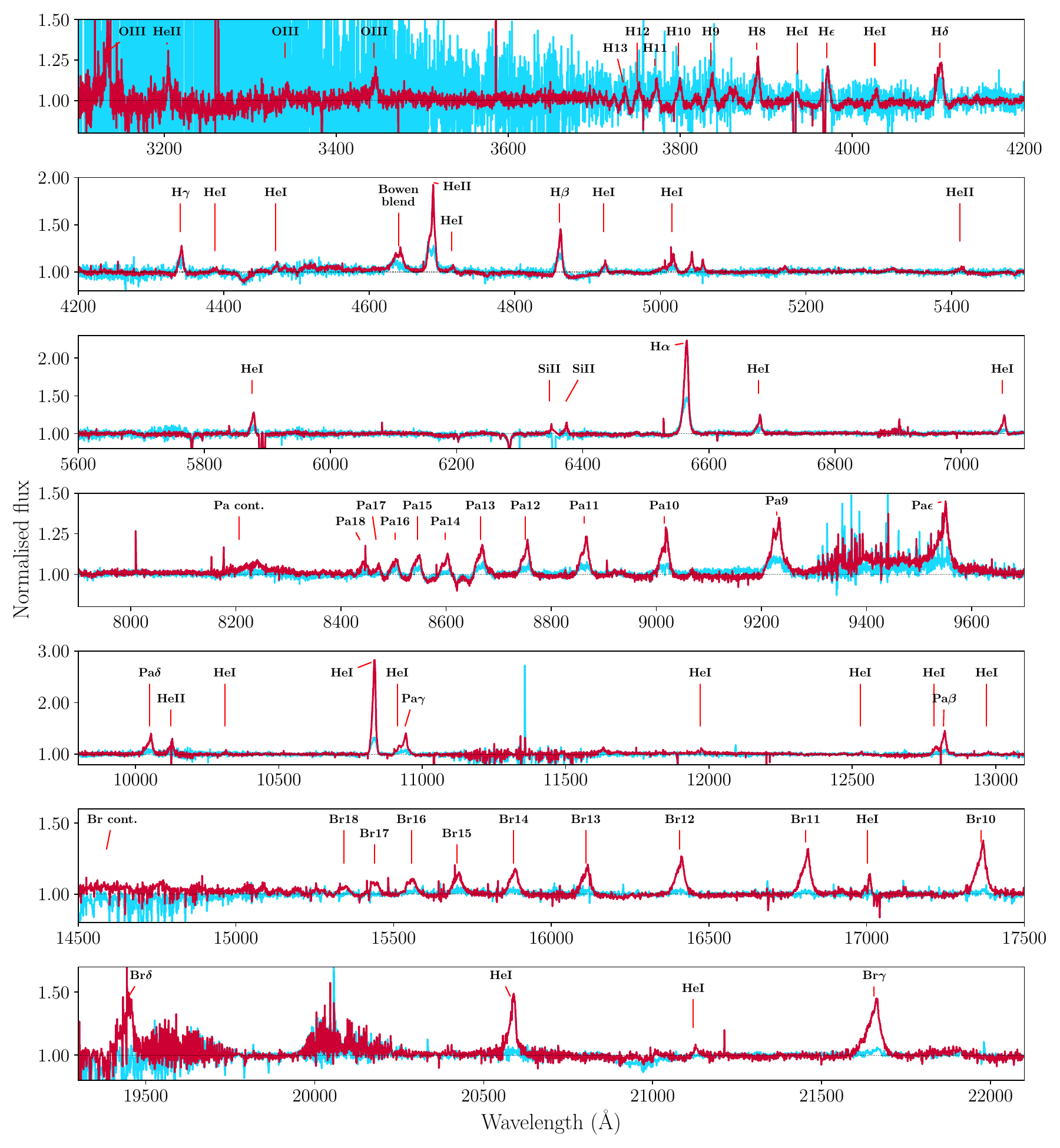}}
    \caption{UVB, VIS and NIR normalised spectra of the soft state S2021a (red) and hard state H2015 (cyan) epochs.}
\label{bigS2021aH2015}%
\end{figure*}

The best-fit spectral indices are reported in Table \ref{table_best_fit_cont}. The errors are primarily driven by uncertainties in interstellar reddening. The SEDs of the soft state epochs are best modeled by a power law with spectral indices of $\rm\alpha = 1.6 \pm 0.3$ and $\rm\alpha = 1.7 \pm 0.3$ for S2021a and S2021b, respectively. A slight deviation from this slope is observed at the red end in the NIR, hinting at a possible NIR excess, which is discussed in Sec. \ref{jet_discussion}. As a test, a broken power-law fit was also applied, yielding consistent values for $\rm\alpha_{\rm{OPT}}$ and $\rm\alpha_{\rm{NIR}}$. Therefore, we prefer to retain the single power-law fit.

In the hard state, the best-fit spectral indices for H2013 are $\rm\alpha_{\rm{OPT}} = 0.1 \pm 0.2$ and $\rm\alpha_{\rm{NIR}} = -0.43 \pm 0.08$. For H2015, we find $\rm\alpha_{\rm{OPT}} = 0.8 \pm 0.3$ and $\rm\alpha_{\rm{NIR}} = -0.08 \pm 0.08$. As expected from visual inspection, the spectral indices in the hard state continua significantly differ from those measured in the soft state, suggesting that distinct radiative processes may be at work (see Sec. \ref{jet_discussion}).

\begin{figure*}
    \centering
  \includegraphics[width=0.499\textwidth]{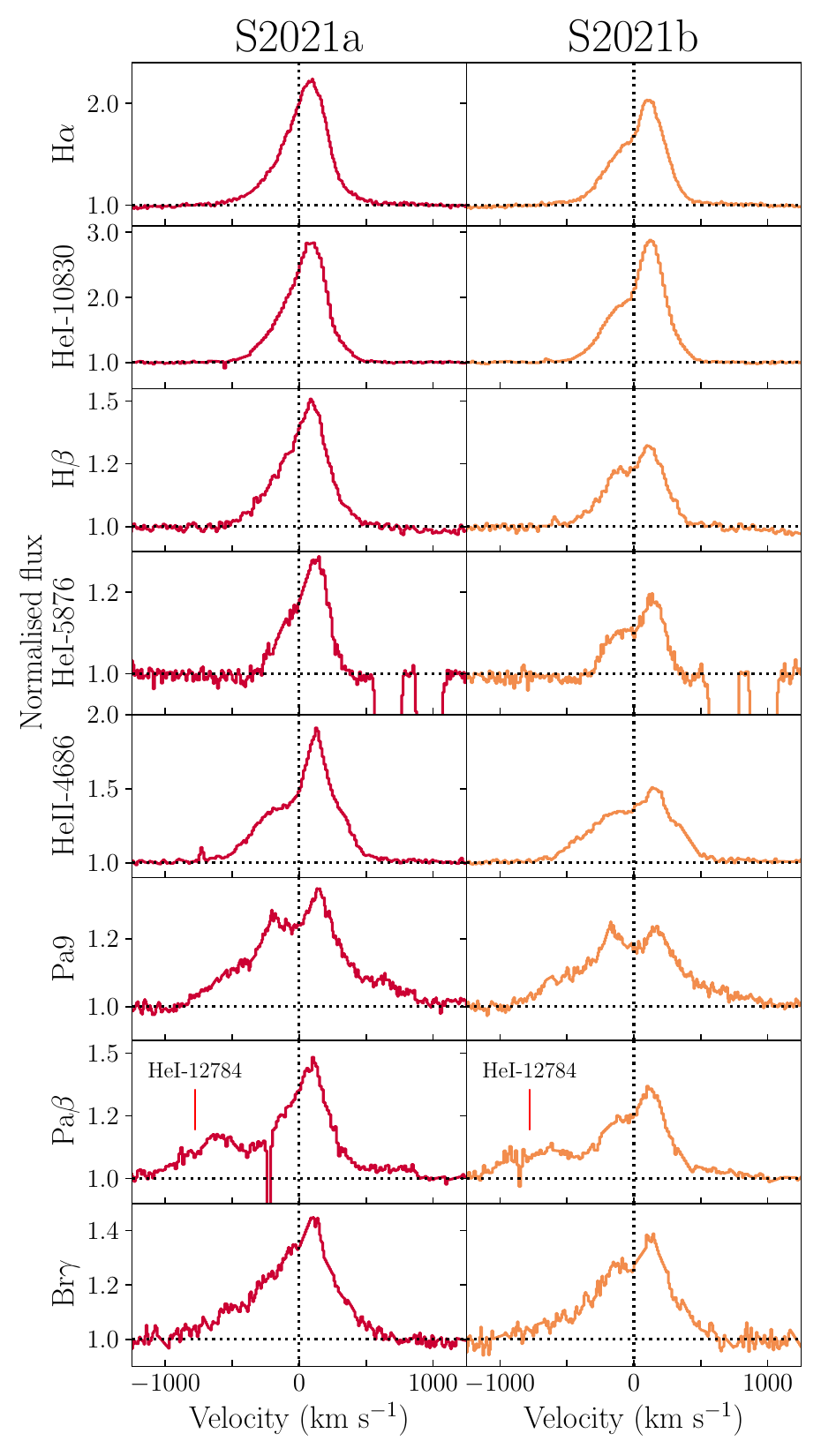}
  \hspace{0\textwidth}  
  \includegraphics[width=0.4829\textwidth]{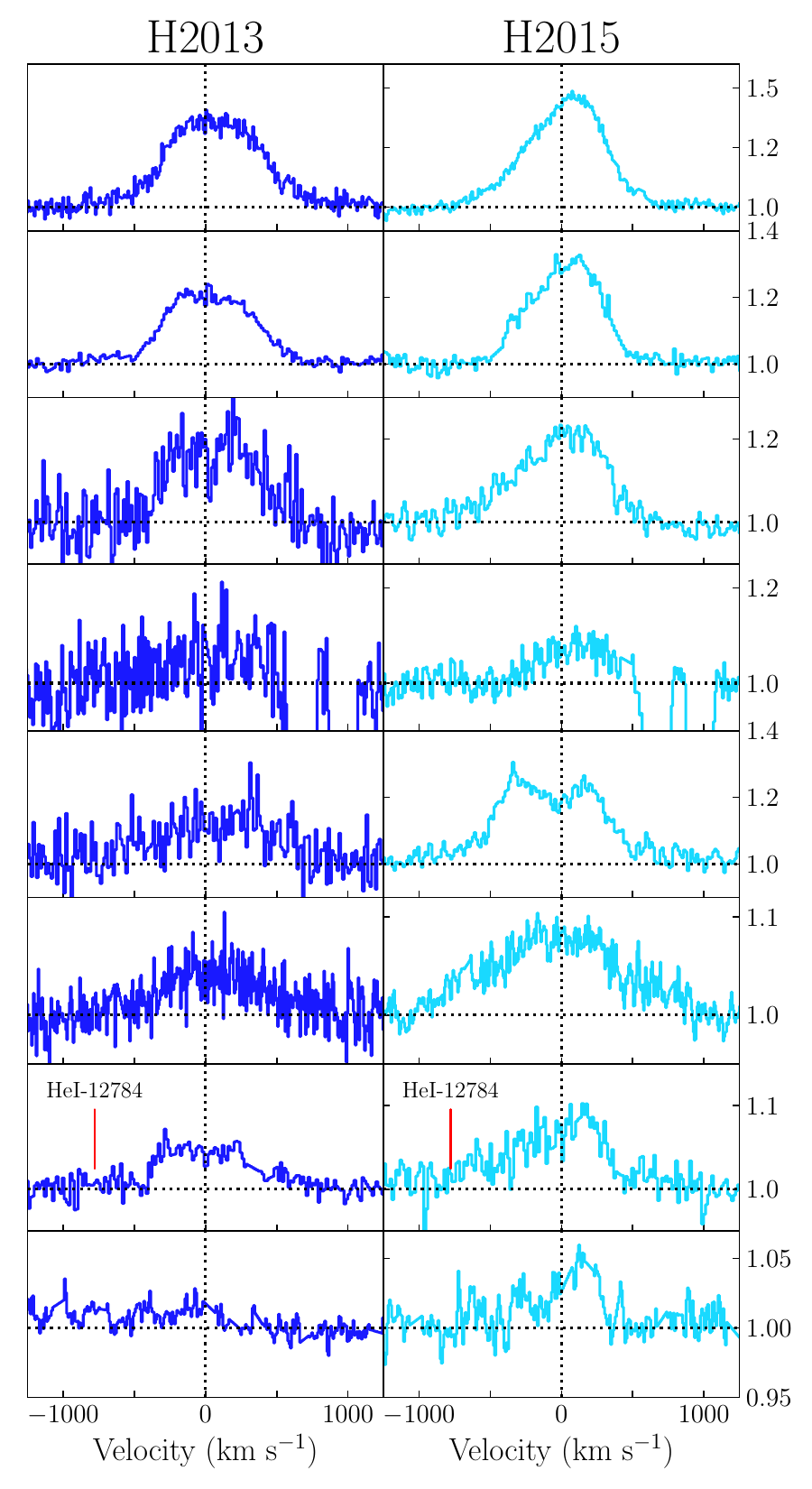}
  \caption{Line profiles of a selection of the most intense features across all four epochs, normalised with respect to their adjacent continuum. The velocities are given in the observer's rest frame.}\label{linesevolution}
\end{figure*} 
    
\subsection{Evolution of the emission lines}
The normalised spectra of all epochs were obtained by fitting a low-order spline to the data after masking the emission lines and strong telluric absorptions. The normalised spectra of the S2021a (red) and H2015 (cyan) epochs are shown in Fig. \ref{bigS2021aH2015}. For completeness, the H2013 and S2021b spectra are also reported in Appendix \ref{appendix}. Several emission features can be observed, with line shapes and intensities varying across the epochs. The soft state spectra appear widely populated, with spectral features corresponding to the Balmer, Paschen and Brackett series up to high orders, \ion{He}{i} and \ion{He}{ii} transitions and also signatures of \ion{O}{iii} and of \ion{O}{ii}, \ion{C}{iii} and \ion{N}{iii} (Bowen blend). Some lines appear particularly intense, such as \ion{H$\rm{\alpha}$}{} and \ion{He}{i}-$10830$. Conversely, only a few emission lines are clearly detected in the hard state, such as some Balmer, Paschen, \ion{He}{i} and \ion{He}{ii} transitions, being \ion{H$\rm{\alpha}$}{}, \ion{H$\rm{\beta}$}{}, \ion{He}{ii}–$4686$ and \ion{He}{i}-$10830$ the most intense features. Fig. \ref{bigS2021aH2015} highlights the relative strength of the lines during the soft state epoch S2021a over those in the hard state (H2015), with the Brackett transitions almost indiscernible from the continuum in the latter case. 

In order to quantify the relative strength of the most prominent emission lines with respect to their underlying continuum, we determined their equivalent width (EW). Features contaminated by adjacent lines, DIBs or strong telluric absorptions were excluded from the calculation. We also computed the V/R ratio (defined as the integrated flux ratio between the blue and red parts of the line, measured from the systemic velocity of the source), which is sensitive to line asymmetries. For completeness, we calculated the full width at zero intensity (FWZI), as this parameter may also be useful in future studies. The results are presented in Table \ref{tableEW}. 

The analysis confirms key features of the emission lines, such as their being stronger during soft state epochs, as evidenced by higher EW values. The most dramatic EW changes occur in the NIR, where emission lines are either undetected or show a substantial drop during the hard state compared to the soft state. The V/R ratio analysis shows that, during soft state epochs, most lines are asymmetric, with the red part dominating over the blue (V/R < 1). A similar trend is observed during the hard state epoch H2013; however, both red- and blue-dominated lines are seen in H2015. The FWZI shows a significant dispersion, ranging from several hundred to $\sim 2000$ km s$^{-1}$, depending on the line and epoch. In particular, the optical hydrogen lines detected during the soft state are narrower than those observed in the hard state or the NIR hydrogen lines in the soft state.

\subsubsection{Soft state: main lines and their properties}

A selection of the most prominent emission lines observed across the epochs is shown in Fig. \ref{linesevolution}. The two columns on the left include features present in the soft state spectra. Although these spectra were collected only four days apart during the same accretion state and with almost identical flux, the emission lines exhibit differences in both intensity and shape.

The low-ionisation features \ion{H$\rm{\alpha}$}{}, \ion{H$\rm{\beta}$}{}, \ion{He}{i}-$5876$ and \ion{He}{i}-$10830$ appear asymmetric and skewed towards redder wavelengths in both epochs. In particular, they are characterised by a red-shifted peak in S2021a, while a shallower blue peak is also present in S2021b. 
During both epochs, \ion{H$\rm{\alpha}$}{} wings meet the continuum at $\sim\pm600$ km s$^{-1}$, while its EW is $\sim10$\AA $ $ (Table \ref{tableEW}).  The V/R ratio is significantly lower than unity for most emission lines, indicating that the blue part of the line is weaker than the red part. This trend is consistently observed in lines that are thought to be good wind tracers, such as H$\rm{\alpha}$, \ion{He}{i}-$6678$ and \ion{He}{i}-$7065$ (see, e.g. \citealt{MunozDarias2019}).

The higher ionisation features, such as the Bowen blend, are strong in both soft state spectra.  \ion{He}{ii}-$4686$ stands out especially in S2021a, with an EW of $\sim6$\AA. In both S2021a and S2021b, the line exhibits a strongly asymmetric profile, with a prominent red peak and a shallower blue peak. 

In the NIR, \ion{H}{i} lines are broader than in the optical, meeting the continuum at $\sim\pm700$ km s$^{-1}$ (compared to $\sim\pm500$ km s$^{-1}$ in the optical). 
\ion{Pa9}{} exhibits a broad double-peaked profile, whose wings meet the continuum at approximately $\pm1000$ km s$^{-1}$.  Double-peaked profiles are common in S2021b (e.g. in \ion{He}{i}-$4922$, \ion{He}{i}-$6678$, \ion{He}{i}-7065, and high-order Paschen and Brackett emission lines), whereas \ion{Pa9}{} is the only double peaked feature in S2021a. Finally, the profiles of \ion{Pa}{$\rm{\beta}$} and \ion{Br$\rm{\gamma}$}{} share common features, including a prominent red-shifted peak. However, \ion{Pa}{$\rm{\beta}$} is contaminated by the adjacent \ion{He}{i}-$12784$ line, preventing a proper characterisation of the line wings. \ion{Br$\rm{\gamma}$}{} exhibits asymmetric wings, the blue one reaching $\sim-1000$ km s$^{-1}$, while the red one meets the continuum at $\sim+700$ km s$^{-1}$ in S2021a.

\subsubsection{Hard state: main lines and their properties}

The relative intensity of the spectral features is lower in the hard state compared to the soft state and they consistently exhibit broader wings (see Table \ref{tableEW}).
\ion{H$\rm{\alpha}$}{} and \ion{He}{i}-$10830$ are the strongest spectral features. In H2013 their profiles appear flat-top, while in H2015 they show a red-skewed asymmetry with a red-shifted peak, as commonly observed in the soft state epochs. As for the higher ionisation features, \ion{He}{ii}-$4686$ emerges as a strong double-peaked line in H2015, with a peak-to-peak separation of $\sim 490$ km s$^{-1}$  and an extended blue wing that reaches $\sim-1000$ km s$^{-1}$. 
In the NIR, \ion{Pa}{9} shows extended wings up to  $\sim1000$ km s$^{-1}$ in  H2015, as observed also in the soft state. During H2013, \ion{Pa}{$\rm{\beta}$} is flat-top, similarly to the most prominent optical lines. Most H and K-band emission lines are significantly weaker than in the soft state or completely absent.

\section{Discussion}

\begin{figure*}
    \centering
  \includegraphics[width=0.49\textwidth]{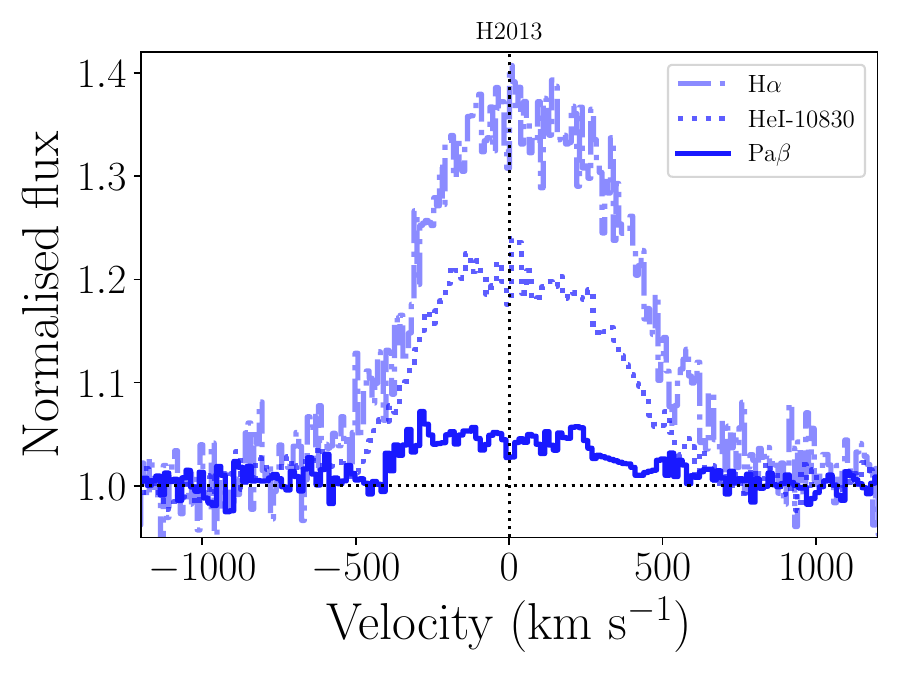}
  \hspace{0.0\textwidth}  
  \includegraphics[width=0.49\textwidth]{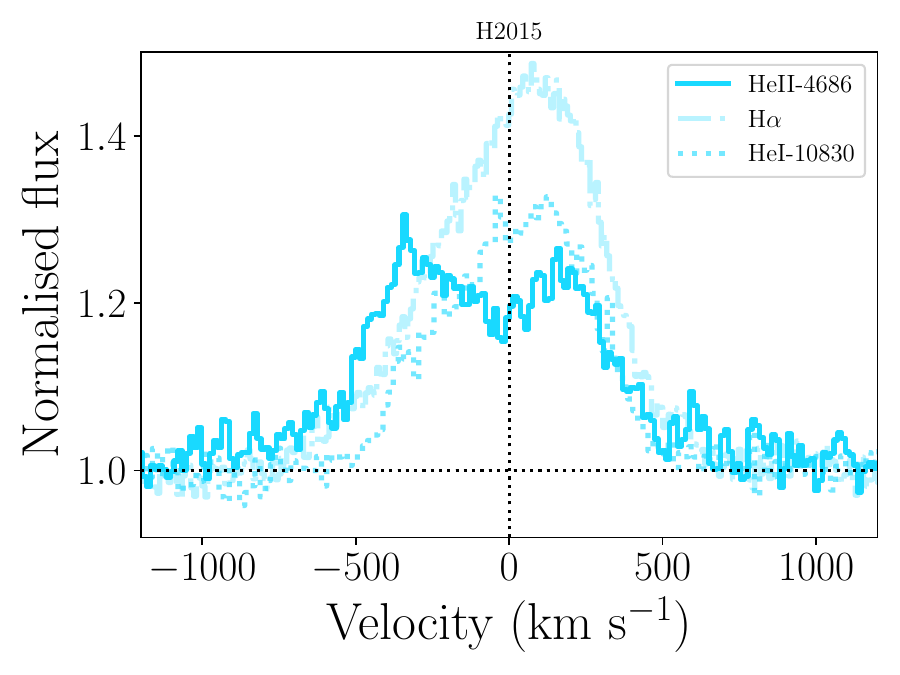}
  \caption{Normalised hard state spectra centred on the most prominent line profiles. The left panel shows the flat-top profiles of H$\rm{\alpha}$, \ion{He}{i}-10830 and \ion{Pa}{$\rm{\beta}$} during epoch H2013. The right panel highlights the difference between the red-skewed, single-peaked profiles of H$\rm{\alpha}$ and \ion{He}{i}-$10830$ and the double peak of \ion{He}{ii}-$4686$ during H2015.}
  \label{importantlines_hardstate}
\end{figure*}

We have presented four epochs of optical and NIR spectroscopy obtained during three different outbursts of GX 339$-$4. Both the continuum and spectral lines exhibit variations across the epochs, with large differences between the hard and the soft states. In the former, the SED shows an enhanced NIR flux, coinciding with the disappearance of most emission lines in the wavelength range corresponding to the H and K bands. During both accretion states, the analysis of the spectral lines does not reveal the presence of P-Cygni profiles. However, we observe line profiles which are often associated with outflows, such as flat-top and asymmetric profiles as well as broad line wings. These signatures are consistent with those observed in other BHTs (e.g. \citealt{Cuneo2020, MataSanchez2024}). 

Some selected recent studies of newly discovered BHTs benefit from a larger number of spectroscopic epochs (e.g. \citealt{MunozDarias2019, Panizo2022}), allowing for statistical tests to compare them (see also \citealt{MataSanchez2023}). Our work, by contrast, includes four epochs taken at different outburst stages of the prototypical system GX~339$-$4. While this limits statistical analyses, the high resolution and sensitivity of our data allowed us to study numerous emission lines across each epoch, revealing consistent patterns in their profiles. These findings further contribute to our understanding of this fundamental system, as well as to the impact that outflows (both jets and winds) have on the optical and NIR spectra of BHTs.

\subsection{The spectral energy distribution}
\label{jet_discussion}
The optical-NIR SED (Fig. \ref{continuum}) exhibits significant variations between the soft and hard state epochs. In the soft state spectra, the best-fit spectral indices are in the range of $1\lesssim\rm{\alpha}\lesssim2$. This is consistent with a thermal origin for the radiation (e.g. \citealt{Hynes2005}). However, in the NIR frequency range, the SED appears to deviate from a simple power law at the red end. A similar behaviour has been observed in at least one other source, the BHT MAXI J1820+070, which shows a comparable infrared excess during a soft state epoch \citep{Koljonen2023}. This was modeled by including the presence of a wind (or an atmosphere) extending above the disc, which could act as a reprocessing site for accretion disc radiation, yielding the infrared excess. Our soft state spectra suggest the presence of wind-type ejecta, and thus, the above interpretation might also apply to GX~339$-$4. However, we caution the reader that the evidence for this excess is relatively weak, as it is found at the red end of our coverage. More data (e.g. mid-infrared) would be needed to properly study and discuss this feature.

The hard state SED is significantly different from that of the soft state. In particular, the NIR continuum is clearly enhanced, which, together with the weakness (H2015) or total absence (H2013) of emission lines in the H and K bands, suggests  the presence of an additional component contributing to these frequency bands. As a matter of fact, the optically thin synchrotron emission generated by the compact jet can be observed up to the optical/infrared range (e.g. \citealt{Corbel2001, Fender2001}) and it can dilute the line profiles, hampering their detection. In the NIR range, its contribution is usually well approximated by a power law with spectral index $-1\leq\rm{\alpha}\leq-0.5$ (e.g. \citealt{Russell2013}). This scenario is supported by the broken power law of $\rm{\alpha_{NIR}}\sim-0.5$ for H2013. In H2015, the flatter spectral index ($\rm{\alpha_{NIR}}\sim-0.1$) might be attributed to the presence of an additional (bluer) component, perhaps associated to the accretion disc.

The contribution of the synchrotron component to the optical-NIR regime (e.g. \citealt{Corbel2002, Shidatsu2011}) and thus the weakening of the emission lines at these wavelengths (\citealt{Rahoui2012}) have been observed in previous outbursts of GX 339$-$4. The thus produced NIR excess has been found to be particularly bright in GX 339$-$4 in the hard state (e.g. \citealt{Corbel2002, Homan2005, Buxton2012}), with a dramatic drop/rise during transitions from/to the hard state (\citealt{Cadolle2011, Dincer2012, Saikia2019}).  As discussed above, the jet component can comfortably account for the observed NIR SED of GX 339$-$4 during the hard state. However, we note that other physical components, such as the so-called hot flow (e.g. \citealt{Kosenkov2020}), might also contribute.

\subsection{Accretion disc winds}
Optical and NIR wind signatures have been detected in several BHTs, although only a few detections have been made in low-to-intermediate inclination systems (e.g. \citealt{Panizo2022}). Our spectroscopy  reveals the presence of flat-top profiles during the hard state epoch H2013 in H$\rm{\alpha}$, \ion{He}{i}-10830 and \ion{Pa}{$\rm{\beta}$} (Fig. \ref{importantlines_hardstate}). Low-velocity
components associated with asymmetric emission from the outer
disc or the companion star can generate these line profiles (e.g. \citealt{Casares2006}).  However, flat-top profiles can also be produced by accretion disc winds, as discussed in theoretical studies (e.g. see \citealt{Murray1995} for a study on active galactic nuclei) and demonstrated using simple toy models (\citealt{MataSanchez2023}). Indeed, flat-top profiles have been previously reported in a number of BHTs (e.g. \citealt{Cuneo2020, Panizo2022, MataSanchez2024}). 

The spectrum H2015 shows single-peaked, red-skewed profiles in H$\rm{\alpha}$ and \ion{He}{i}-$10830$. Line asymmetries can be attributed to asymmetric emission from the accretion disc (i.e. a hot spot; \citealt{Stover1981}). However, the double-peaked shape of \ion{He}{ii}-$4686$ (Fig. \ref{importantlines_hardstate}) and high order Paschen lines are consistent with arising in a Keplerian and symmetric accretion disc, arguing against a geometrical explanation for the remaining asymmetric profiles.  Therefore, we propose that these single-peaked lines might be generated in a wind. This phenomenon naturally produces blue-shifted absorptions that might overlap with the disc line profile, resulting in asymmetric, red-skewed shapes. 

Figure \ref{importantlines_hardstate} illustrates the differences between the emission line profiles during the two hard state epochs. These differences are not entirely surprising, as variability in the properties of wind signatures on timescales ranging from minutes to days has been observed in other sources (e.g. \citealt{MataSanchez2018, MunozDarias2019}). Such variability is perhaps more expected when comparing two epochs from different outbursts, even though they occur in a similar hard state. Different line properties are thought to result from variations in the ionisation and density conditions of the wind, which can change due to several factors, such as irradiation from the central source and inhomogeneities along the line of sight.

During the soft state epochs, broad line wings are observed in intense NIR lines (i.e. \ion{Pa9}{}, \ion{Br11}{}, \ion{Br10}{}, \ion{Br$\rm{\gamma}$}{}). Broad line wings are a common feature in BHTs (e.g. \citealt{Munoz2016, Munoz2018, MunozDarias2019, Sanchezsierras23_1915, Panizo2022}) and can be produced by an expanding envelope. Additionally, most of these lines are asymmetric.  Again, the presence of a hot spot could be invoked to explain this feature, however, we observe that asymmetric, red-skewed profiles coexist with double-peaked lines such as \ion{Pa9}{} (Fig. \ref{linesevolution}), weakening a geometric explanation. The orbital phase difference between the two soft state epochs is $\Delta\Phi = 0.23\pm0.01$. The presence of similar line profiles with an enhanced red peak at phases separated by a quarter of the orbital period argues against a hot spot origin, unless it is significantly extended in orbital phase. Unfortunately, we cannot calculate the relative orbital phases for the hard state epochs due to large uncertainties in the projected orbital solution (\citealt{Heida2017}). Nevertheless, the fact that red-skewed asymmetries are also observed in H2015 (Fig. \ref{linesevolution}), coupled with the absence of any instances of blue-skewed profiles in the four epochs analysed in this study,  further argues against a phase-dependent origin for this feature.

In the optical range, wind signatures are limited to the presence of red-skewed, asymmetric profiles. Although a geometrical explanation cannot be fully ruled out, it could hardly account for the simultaneous observation of broad line wings in the NIR. Conversely, the wind scenario has the potential to simultaneously explain the observed features, including broad line wings, red-skewed asymmetry, and flat-top profiles across all inspected accretion states. 

The possible presence of outflow signatures in the optical spectrum of GX 339$-$4 was explored by \citeauthor{Soria1999} (\citeyear{Soria1999}) and \citeauthor{Wu2001} (\citeyear{Wu2001}), who identified single-peaked and asymmetric \ion{H$\rm{\alpha}$}{} lines, simultaneous with double-peaked \ion{He}{ii}-4686 profiles. In these works, the single-peaked lines were interpreted as a hint for the presence of an outflow, either a dense wind originated in the disc or an evaporating corona. Broad \ion{Pa$\rm{\beta}$}{} wings were also identified by \cite{Rahoui2014}, who suggested a wind origin to explain them. The higher resolution and broad coverage of our data allowed for a detailed analysis of multiple emission lines, revealing wind signatures across both the optical and NIR spectra in both hard and soft states. Thus, the spectral features identified in this study are consistent with those observed previously in this system and in other low-to-intermediate inclination BHTs (see also e.g. \citealt{Panizo2022}).

\subsection{The role of orbital inclination and ionisation state in the wind
detectability}\label{inclin}

According to the current paradigm, which describes the geometry of accretion disc winds in black hole transients as equatorial (e.g. \citealt{Ponti2012}), the detection of P-Cygni profiles in low-to-intermediate inclination systems may be challenging. The reason being that the blue-shifted absorption component caused by the wind would occur at lower velocities, due to projection effects  (e.g. \citealt{Panizo2022}; see also \citealt{Higginbottom2019} for a theoretical study). This can make the (wind-related) absorption feature overlap with the (disc) emission line profile, leading to the suppression of the blue peak, either completely (producing skewed profiles, e.g.  \citealt{MunozDarias2019, Panizo2022}),  or partially (leaving a red bump, e.g. \citealt{MunozDarias2020}). The predominant features associated with outflows in these systems would instead consist of broad emission wings, asymmetric red-skewed profiles, and low-velocity blue-shifted absorptions. Taking this into account, the potential outflow signatures that we have identified in GX 339$-$4 are in agreement with the intermediate inclination proposed for this system (\citealt{Zdziarski2019}).

During the hard state of GX 339$-$4, wind-related features are only detected in low-ionisation transitions (e.g. \ion{H$\rm{\alpha}$}{}, \ion{He}{i}-10830), in agreement with observations of other BHTs. Only a few systems show  optical wind signatures in high ionisation lines (e.g. \citealt{Charles2019, Jimenez2019, MataSanchez2024}). These are interpreted as indicative of a hot and dense outflow. During the soft state epochs, we observed an enhanced red peak in both low ionisation lines and \ion{He}{ii}-$4686$. Assuming these features are produced by a wind, they suggest changes in the physical conditions of the ejecta during different accretion states, being more highly ionised and denser during the soft state.

\begin{figure}
    \centering
    \includegraphics[width=0.5\textwidth]{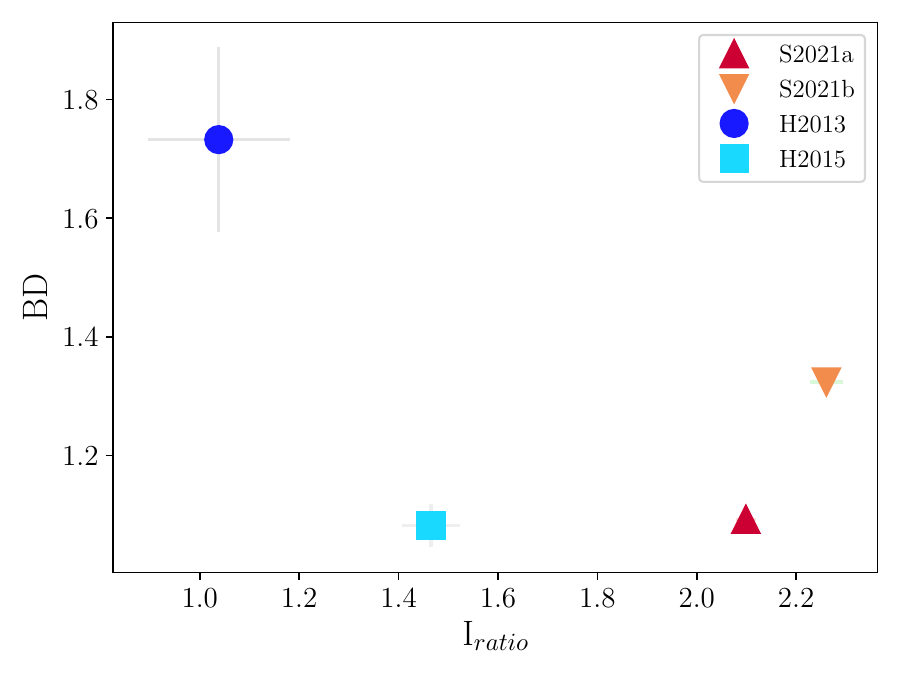}
    \caption{Balmer decrement (BD) against the ionisation ratio (I$_{\rm{ratio}}$) for each observing epoch.}
    \label{BD}%
\end{figure}

The flux ratio I$_{\rm{ratio}}=$ \ion{He}{ii}-$4686/$\ion{H$\rm{\beta}$}{} is a good tracer of the ionisation state (e.g. \citealt{Munoz2016}) and was calculated for each epoch (see Fig. \ref{BD}). We find  I$_{\rm{ratio}}>2$ for the soft state spectra, whereas I$_{\rm{ratio}}\sim 1-1.5$ is found during H2013 and H2015. This is consistent with more highly ionised material during the soft state epochs. The Balmer ecrement (BD), which is defined as the flux ratio between \ion{H$\rm{\alpha}$}{} and \ion{H$\rm{\beta}$}{}, is a proxy for optically thinner gas (e.g. nebular phases, \citealt{Munoz2016}). The BD of each epoch has been calculated, and its values reported in Fig. \ref{BD}. We find BD $\sim1-1.5$ in both accretion states, which is roughly consistent with previous reports (e.g. \citealt{Rahoui2014}). Such low BD values are typically associated with emission lines originating in dense regions, e.g. the accretion disc. We do not find evidence for nebular phases, characterised by a BD $\gtrsim 3$ (e.g. \citealt{MataSanchez2018}).

To summarise, we consistently observe spectral features classically associated with the presence of winds across the four optical-to-NIR spectroscopic epochs analysed in this work. Broad emission line wings (e.g. Pa9, Br$\rm{\gamma}$) reaching velocities up to $\sim1000$ km s$^{-1}$ are detected, compatible with the typical velocities of a few thousand km s$^{-1}$ commonly observed in BHT winds (e.g. \citealt{Munoz2018, MataSanchez2024}). The red-skewed and flat-top line profiles observed also agree with the cold wind signatures detected in MAXI~J1348–630, arguably the best example of winds observed so far in a low- to intermediate-inclination BHT \citep{Panizo2022}.

The detection of wind signatures across several emission lines during all the spectroscopic epochs analysed here, combined with the profiles of \ion{H$\rm{\alpha}$}{} and \ion{Pa$\rm{\beta}$}{} reported in previous studies \citep{Soria1999, Wu2001, Rahoui2014}, supports the presence of wind-type outflows in GX~339$-$4. In this context, the absence of P-Cygni line profiles in the spectra can be attributed to the orbital inclination of the system, which is thought to be lower than that of the sources showing these features.

\section{Conclusions}
We have analysed four epochs of simultaneous optical and infrared spectroscopy of the black hole transient GX 339$-$4 with emphasis on searching for outflow-related features. We do not detect P-Cygni line profiles as those observed in other systems of this class. However, we detect other potential wind signatures, such as asymmetric and flat-top line profiles, as well as broad emission line wings. During the hard state, the wind signatures are mainly observed in the optical. Additionally, signatures of the jet are evident in the form of an enhanced NIR continuum, coinciding with the disappearance of most H and K bands' emission lines. During the soft state, the strongest wind signatures are observed in the NIR spectral range. We discuss that the absence of P-Cygni profiles can be attributed to the moderately low inclination of the system. Additional optical and NIR spectroscopy of GX 339$-$4 and other low-to-intermediate inclination systems in outburst is encouraged to delve deeper into the optical and infrared observational properties of black hole transients.

\begin{acknowledgements}
We acknowledge support by Spanish \textit{Agencia estatal de investigaci\'on} via PID2020-120323GB-I00 and PID2021-124879NB-I00. Based on observations collected at the European Southern Observatory under ESO programmes 105.200K.001, 105.200K.002, 291.D-5047(A) and 295.D-5027(C). This work made use of data supplied by the UK Swift Science Data Centre at the University of Leicester.
M.A.P. acknowledges support through the Ramón y Cajal grant RYC2022-035388-I, funded by MCIU/AEI/10.13039/501100011033 and FSE+. This material is based upon work supported by Tamkeen under the NYU Abu Dhabi Research Institute grant CASS. GP acknowledge financial support from the European Research Council (ERC) under the European Union's Horizon 2020 research and innovation program HotMilk (grant agreement No. 865637). GP acknowledges support from Bando per il Finanziamento della Ricerca Fondamentale 2022 dell'Istituto Nazionale di Astrofisica (INAF): GO Large program and from the Framework per l'Attrazione e il Rafforzamento delle Eccellenze (FARE) per la ricerca in Italia (R20L5S39T9).  We acknowledge Astropy,\footnote{\url{http://www.astropy.org}} a community-developed core Python package for Astronomy \citep{astropy2013, astropy2018}.
\end{acknowledgements}

   \bibliographystyle{aa} % style aa.bst
   %\bibliography{biblio} % your references 
   
 % your references Yourfile.bib

\begin{appendix} 

\begin{figure*}
\section{H2013 and S2021b Spectra}\label{appendix}
        \centering
        \includegraphics[width=\textwidth]{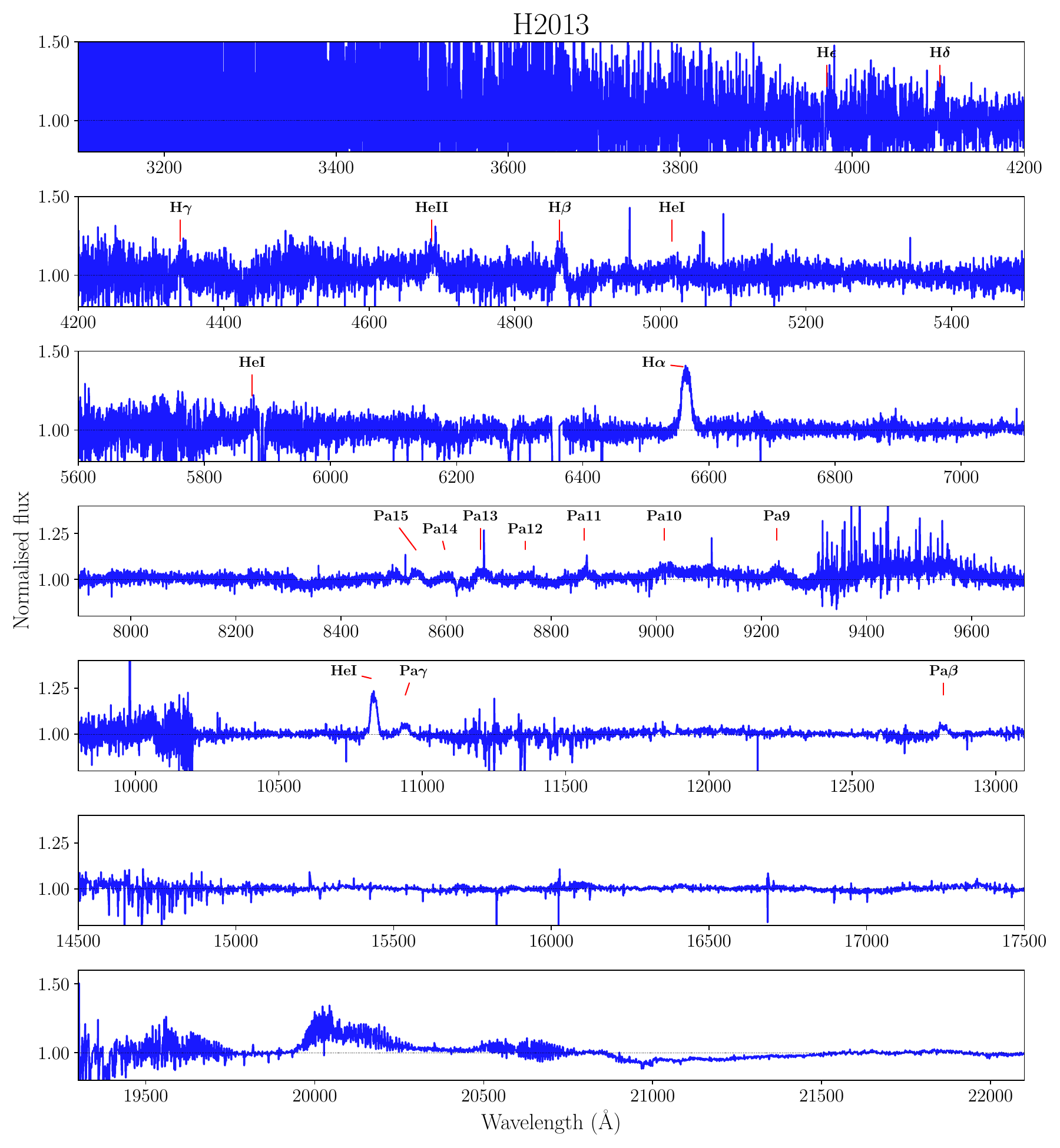}
        \caption{UVB, VIS, NIR spectrum of the epoch H2013.}
        
\end{figure*}

\begin{figure*}
    \centering
    \includegraphics[width=\textwidth]{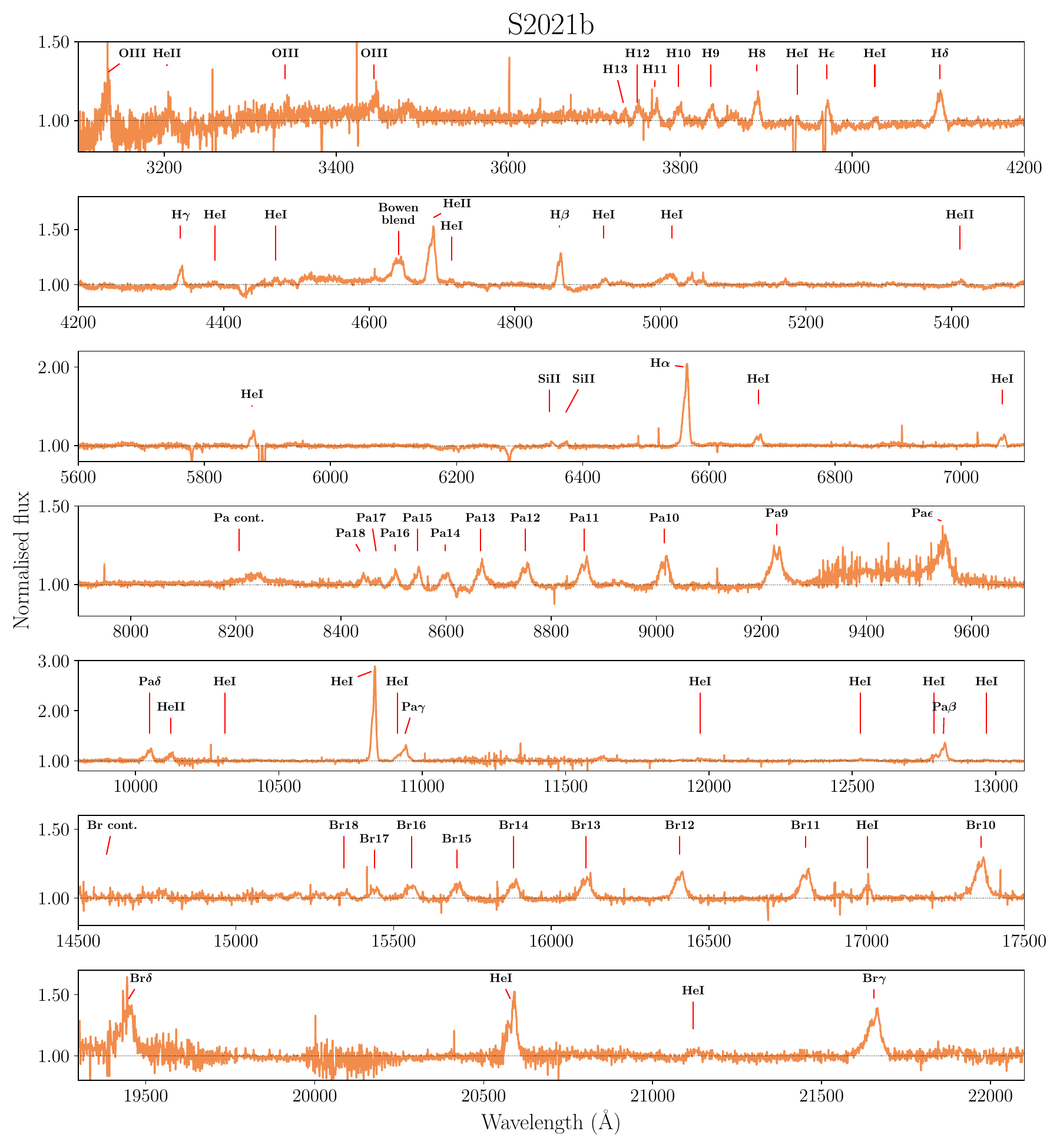}
    \caption{UVB, VIS, NIR spectrum of the epoch S2021b.}
    \label{}%
\end{figure*}

\begin{center}
\onecolumn
\section{Table}\label{EWtable}
\begin{longtable}{|c|c|ccc|}
\caption{EW, V/R ratio and FWZI of the most prominent emission lines.}          
\label{tableEW}   
\\
\hline                      
Epoch & Line & EW (\AA) & V/R & FWZI (km s$^{-1}$) \\ 
\hline  
 S2021a  & \ion{OIII-3444}{} & $0.94\pm0.06$ & $0.63\pm0.1$ & $1350\pm140$ \\ 
 &  \ion{H10}{}  & $0.77\pm0.03$ & $0.46\pm0.05$ & $1181\pm53$ \\ 
 & \ion{H9}{} & $1.06\pm0.03$ & $0.58\pm0.04$ & $1184\pm55$ \\ 
 & \ion{H8}{} & $1.5\pm0.03$ & $0.71\pm0.04$ & $1179\pm45$ \\ 
 & \ion{HeI-4026}{} & $0.26\pm0.02$ & $0.42\pm0.11$ & $1026\pm56$ \\ 
 & \ion{H$\rm{\delta}$}{} & $1.7\pm0.03$ & $1.0\pm0.04$ & $1429\pm33$ \\ 
 & \ion{H$\rm{\gamma}$}{} & $1.69\pm0.03$ & $0.66\pm0.03$ & $1147\pm35$ \\ 
 & \ion{HeI-4388}{} & $0.34\pm0.02$ & $1.46\pm0.24$ & $1470\pm160$ \\ 
 & \ion{HeII-4686}{} & $6.08\pm0.03$ & $0.69\pm0.01$ & $1327\pm10$ \\ 
 & \ion{HeI-4713}{} & $0.35\pm0.02$ & $0.66\pm0.11$ & $946\pm62$ \\ 
 & \ion{H$\rm{\beta}$}{} & $3.17\pm0.03$ & $0.76\pm0.01$ & $1170\pm18$ \\ 
 & \ion{HeI-4922}{} & $0.79\pm0.02$ & $0.51\pm0.03$ & $996\pm36$ \\ 
 & \ion{HeII-5411}{} & $0.73\pm0.02$ & $0.82\pm0.06$ & $1444\pm87$ \\ 
 & \ion{HeI-5876}{} & $1.72\pm0.02$ & $0.52\pm0.02$ & $728\pm22$ \\ 
 & \ion{SiII-6371}{} & $0.68\pm0.03$ & $0.3\pm0.04$ & $1040\pm79$ \\ 
 & \ion{H$\rm{\alpha}$}{} & $10.51\pm0.04$ & $0.73\pm0.01$ & $1420\pm27$ \\ 
 & \ion{HeI-6678}{} & $1.64\pm0.02$ & $0.51\pm0.02$ & $935\pm23$ \\ 
 & \ion{HeI-7065}{} & $1.36\pm0.01$ & $0.43\pm0.01$ & $687\pm17$ \\ 
 & \ion{HeI-7281}{} & $0.65\pm0.02$ & $0.29\pm0.03$ & $685\pm36$ \\ 
 & \ion{OI-7774}{} & $0.58\pm0.02$ & $0.64\pm0.05$ & $872\pm49$ \\ 
 & \ion{Pa12}{} & $2.97\pm0.02$ & $0.7\pm0.01$ & $1284\pm17$ \\ 
 & \ion{Pa11}{} & $3.62\pm0.04$ & $0.81\pm0.02$ & $1409\pm23$ \\ 
 & \ion{Pa10}{} & $4.35\pm0.04$ & $0.82\pm0.02$ & $1488\pm23$ \\ 
 & \ion{Pa9}{} & $7.63\pm0.05$ & $0.95\pm0.01$ & $2091\pm24$ \\ 
 & \ion{HeII-10124}{} & $4.45\pm0.06$ & $0.94\pm0.03$ & $1396\pm43$ \\ 
 & \ion{HeI-10311}{} & $0.61\pm0.13$ & $0.34\pm0.25$ & $1110\pm260$ \\ 
 & \ion{HeI-10830}{} & $24.81\pm0.48$ & $0.7\pm0.01$ & $957\pm17$ \\ 
 & \ion{HeI-12528}{} & $0.59\pm0.05$ & $0.49\pm0.12$ & $538\pm57$ \\ 
 & \ion{HeI-12968}{} & $0.81\pm0.07$ & $0.5\pm0.11$ & $956\pm83$ \\ 
 & \ion{Br18}{} & $1.36\pm0.11$ & $0.74\pm0.15$ & $1320\pm110$ \\ 
 & \ion{Br17}{} & $2.4\pm0.1$ & $0.92\pm0.09$ & $1150\pm75$ \\ 
 & \ion{Br16}{} & $3.99\pm0.13$ & $0.99\pm0.07$ & $1860\pm150$ \\ 
 & \ion{Br15}{} & $5.33\pm0.15$ & $0.79\pm0.04$ & $1610\pm97$ \\ 
 & \ion{Br14}{} & $5.54\pm0.13$ & $0.83\pm0.04$ & $1585\pm90$ \\ 
 & \ion{Br13}{} & $6.63\pm0.11$ & $0.92\pm0.04$ & $1624\pm74$ \\ 
 & \ion{Br12}{} & $9.06\pm0.08$ & $0.9\pm0.02$ & $1828\pm49$ \\ 
 & \ion{Br11}{} & $9.58\pm0.1$ & $0.86\pm0.02$ & $1703\pm47$ \\ 
 & \ion{Br10}{} & $12.64\pm0.11$ & $0.9\pm0.02$ & $1654\pm25$ \\ 
 & \ion{HeI-17002}{} & $1.84\pm0.06$ & $0.46\pm0.05$ & $605\pm53$ \\ 
 & \ion{Br$\rm{\delta}$}{} & $15.81\pm0.84$ & $0.69\pm0.06$ & $1950\pm150$ \\ 
 & \ion{HeI-20587}{} & $12.14\pm0.14$ & $0.66\pm0.03$ & $1166\pm33$ \\ 
 & \ion{HeI-21126}{} & $1.21\pm0.12$ & $0.16\pm0.11$ & $607\pm65$ \\ 
 & \ion{Br$\rm{\gamma}$}{} & $18.69\pm0.18$ & $1.13\pm0.03$ & $1949\pm30$ \\ 
\hline
S2021b  &  \ion{H10}{}  & $0.55\pm0.03$ & $0.69\pm0.09$ & $1393\pm68$ \\ 
 & \ion{H9}{} & $0.73\pm0.03$ & $0.61\pm0.07$ & $1283\pm68$ \\ 
 & \ion{H8}{} & $1.29\pm0.03$ & $0.87\pm0.05$ & $1335\pm51$ \\ 
 & \ion{HeI-4026}{} & $0.2\pm0.02$ & $0.82\pm0.25$ & $1190\pm120$ \\ 
 & \ion{H$\rm{\delta}$}{} & $1.78\pm0.03$ & $1.14\pm0.04$ & $1740\pm46$ \\ 
 & \ion{H$\rm{\gamma}$}{} & $1.27\pm0.03$ & $0.77\pm0.04$ & $1258\pm46$ \\ 
 & \ion{HeII-4686}{} & $4.8\pm0.03$ & $0.93\pm0.01$ & $1538\pm16$ \\ 
 & \ion{H$\rm{\beta}$}{} & $2.3\pm0.03$ & $0.87\pm0.02$ & $1120\pm27$ \\ 
 & \ion{HeI-4922}{} & $0.55\pm0.02$ & $0.61\pm0.06$ & $1179\pm76$ \\ 
 & \ion{HeII-5411}{} & $0.81\pm0.03$ & $0.98\pm0.07$ & $2070\pm150$ \\ 
 & \ion{HeI-5876}{} & $1.25\pm0.02$ & $0.67\pm0.03$ & $774\pm27$ \\ 
 & \ion{SiII-6371}{} & $0.41\pm0.03$ & $0.7\pm0.11$ & $1020\pm140$ \\ 
 & \ion{H$\rm{\alpha}$}{} & $8.83\pm0.04$ & $0.67\pm0.01$ & $1094\pm20$ \\ 
 & \ion{HeI-6678}{} & $1.27\pm0.02$ & $0.83\pm0.03$ & $918\pm23$ \\ 
 & \ion{HeI-7065}{} & $1.0\pm0.04$ & $0.82\pm0.05$ & $672\pm43$ \\ 
 \caption{continued}\\
 & \ion{HeI-7281}{} & $0.39\pm0.02$ & $0.64\pm0.07$ & $637\pm50$ \\ 
 & \ion{Pa12}{} & $2.48\pm0.04$ & $0.84\pm0.03$ & $1448\pm35$ \\ 
 & \ion{Pa11}{} & $3.03\pm0.04$ & $1.01\pm0.03$ & $1542\pm31$ \\ 
 & \ion{Pa10}{} & $3.58\pm0.04$ & $1.0\pm0.02$ & $1646\pm26$ \\ 
 & \ion{Pa9}{} & $5.94\pm0.04$ & $1.12\pm0.02$ & $2163\pm25$ \\ 
 & \ion{HeII-10124}{} & $3.72\pm0.08$ & $1.2\pm0.06$ & $1823\pm77$ \\ 
 & \ion{HeI-10830}{} & $25.52\pm0.13$ & $0.61\pm0.01$ & $953\pm5$ \\ 
 & \ion{Br16}{} & $4.53\pm0.12$ & $1.16\pm0.08$ & $1860\pm120$ \\ 
 & \ion{Br15}{} & $4.46\pm0.12$ & $0.79\pm0.09$ & $1740\pm110$ \\ 
 & \ion{Br14}{} & $4.83\pm0.11$ & $0.85\pm0.06$ & $1855\pm96$ \\ 
 & \ion{Br13}{} & $6.34\pm0.09$ & $0.92\pm0.04$ & $1600\pm55$ \\ 
 & \ion{Br12}{} & $7.33\pm0.09$ & $0.95\pm0.03$ & $1923\pm57$ \\ 
 & \ion{Br11}{} & $7.75\pm0.08$ & $1.09\pm0.03$ & $1771\pm48$ \\ 
 & \ion{Br10}{} & $12.11\pm0.11$ & $1.09\pm0.03$ & $1698\pm27$ \\ 
 & \ion{HeI-17002}{} & $1.99\pm0.05$ & $0.89\pm0.08$ & $726\pm57$ \\ 
 & \ion{Br$\rm{\delta}$}{} & $15.66\pm0.59$ & $1.18\pm0.07$ & $2050\pm120$ \\ 
 & \ion{HeI-20587}{} & $12.39\pm0.16$ & $0.62\pm0.03$ & $987\pm19$ \\ 
 & \ion{Br$\rm{\gamma}$}{} & $16.0\pm0.25$ & $1.13\pm0.05$ & $1715\pm42$ \\ 
\hline
 H2013 & \ion{HeII-4686}{} & $2.25\pm0.17$ & $0.82\pm0.18$ & $2110\pm260$ \\ 
 & \ion{H$\rm{\beta}$}{} & $2.26\pm0.16$ & $0.81\pm0.15$ & $1610\pm130$ \\ 
 & \ion{H$\rm{\alpha}$}{} & $6.58\pm0.16$ & $0.83\pm0.04$ & $1830\pm79$ \\ 
 & \ion{HeI-10830}{} & $5.81\pm0.12$ & $0.99\pm0.04$ & $1657\pm43$ \\ 
\hline
H2015  & \ion{H$\rm{\delta}$}{} & $0.85\pm0.1$ & $2.07\pm0.78$ & $1710\pm160$ \\ 
 & \ion{H$\rm{\gamma}$}{} & $1.5\pm0.1$ & $1.14\pm0.17$ & $1460\pm140$ \\ 
 & \ion{HeII-4686}{} & $3.46\pm0.06$ & $1.44\pm0.07$ & $1830\pm49$ \\ 
 & \ion{H$\rm{\beta}$}{} & $2.51\pm0.07$ & $1.18\pm0.08$ & $1766\pm71$ \\ 
 & \ion{H$\rm{\alpha}$}{} & $6.55\pm0.06$ & $0.92\pm0.02$ & $1636\pm34$ \\ 
 & \ion{HeI-6678}{} & $0.71\pm0.03$ & $0.83\pm0.12$ & $1182\pm80$ \\ 
 & \ion{HeI-7065}{} & $0.44\pm0.03$ & $0.43\pm0.09$ & $1031\pm75$ \\ 
 & \ion{Pa12}{} & $0.91\pm0.03$ & $0.82\pm0.07$ & $1580\pm58$ \\ 
 & \ion{Pa11}{} & $1.02\pm0.03$ & $1.15\pm0.09$ & $1412\pm60$ \\ 
 & \ion{Pa10}{} & $1.14\pm0.03$ & $1.04\pm0.1$ & $1667\pm68$ \\ 
 & \ion{Pa9}{} & $2.82\pm0.06$ & $1.32\pm0.07$ & $2686\pm93$ \\ 
 & \ion{HeI-10830}{} & $6.66\pm0.13$ & $0.88\pm0.06$ & $1362\pm40$ \\ 
 & \ion{Br$\rm{\gamma}$}{} & $1.46\pm0.07$ & $0.62\pm0.08$ & $950\pm86$ \\ 
\hline          
\end{longtable}
\end{center}

\end{appendix}

\end{document}